\documentclass[%
 reprint,
 amsmath,amssymb,
 aps,soul,
prl,
]{revtex4-2}

\usepackage{graphicx}
\usepackage{dcolumn}
\usepackage{bm}
\usepackage{xcolor}
\usepackage{url}

\usepackage{lipsum}
\makeatletter
\newcommand*{\balancecolsandclearpage}{%
  \close@column@grid
  \cleardoublepage
  \twocolumngrid
}

\renewcommand{\selectlanguage}[1]{}

\begin{document}

\preprint{APS/123-QED}

\title{Characteristic E-region Plasma Signature of \\ Magnetospheric Wave-Particle Interactions}

\author{Magnus F Ivarsen}
\altaffiliation[Now at ]{Department of Physics and Engineering Physics, University of Saskatchewan, Saskatoon, Canada}
\affiliation{Department of Physics, University of Oslo, Oslo, Norway}%

\author{Yukinaga Miyashita}
\altaffiliation[Also at ]{Department of Astronomy and Space Science, Korea University of Science and Technology, Daejeon, South Korea}
\affiliation{Space Science Division, Korea Astronomy and Space Science Institute, Daejeon, South Korea}%

\author{Jean-Pierre St-Maurice}
\altaffiliation[Also at ]{Department of Physics and Astronomy, University of Western Ontario, London, Canada}
\author{Glenn C Hussey}
\author{Brian Pitzel}
\author{Draven Galeschuk}
\author{Saif Marei}
\affiliation{Department of Physics and Engineering Physics, University of Saskatchewan, Saskatoon, Canada}

\author{Richard B. Horne}
\affiliation{British Antarctic Survey, Cambridge, UK}

\author{Yoshiya Kasahara}
\author{Shoya Matsuda}
\affiliation{Graduate School of Natural Science and Technology, Kanazawa University, Kanazawa, Japan}

\author{Satoshi Kasahara}
\author{Kunihiro Keika}
\affiliation{Department of Earth and Planetary Science, University of Tokyo, Tokyo, Japan}

\author{Yoshizumi Miyoshi}
\author{Kazuhiro Yamamoto}
\author{Atsuki Shinbori}
\affiliation{Institute for Space-Earth Environmental Research, Nagoya University, Nagoya, Japan}

\author{Devin R Huyghebaert}
\altaffiliation[Also at ]{Department of Physics and Engineering Physics, University of Saskatchewan, Saskatoon, Canada}
\affiliation{Department of Physics and Technology, University of Tromsø, Tromsø, Norway}%

\author{Ayako Matsuoka}
\affiliation{Data Analysis Center for Geomagnetism and Space Magnetism, Kyoto University, Kyoto, Japan}

\author{Shoichiro Yokota}
\affiliation{Department of Earth and Space Science, Osaka University, Toyonaka, Japan}

\author{Fuminori Tsuchiya}
\affiliation{Planetary Plasma and Atmospheric Research Center, Tohoku University, Tohoku, Japan}


\begin{abstract}
Plasma waves in the magnetosphere scatter electrons, causing them to precipitate into Earth's atmosphere, imparting their temporal characteristics to diffuse aurorae. In a case study of conjugate radar- and satellite-observations, we demonstrate a close and unprecedented association between enhanced electrostatic cyclotron harmonic wave activity in the magnetosphere and the appearance of meter-scale plasma turbulence a few seconds later in the lower ionosphere on nearby magnetic field lines. Such \textit{direct} structuring of the ionosphere carries implications for our understanding of space weather.
\end{abstract}

\maketitle

\section{\label{sec:intro}Introduction}

During major disturbances in geospace much energy is imparted  from the solar wind to Earth's atmosphere. Large-scale electrical currents and the creation of strong electric fields at ionospheric altitudes ensues \cite{iijimaLargescaleCharacteristicsFieldaligned1978,keilingAssessingGlobalAlfven2019}. The effect is widespread plasma turbulence \cite{hubaIonosphericTurbulenceInterchange1985,wiltbergerEffectsElectrojetTurbulence2017}.

The Farley-Buneman (FB) instability features prominently in the ionosphere's lower layers, the E-region  \cite{farleyPlasmaInstabilityResulting1963, bunemanExcitationFieldAligned1963}. The intense meter-scale FB structures are excited when the relative drift (polarization electric field) between the strongly magnetized electrons and largely unmagnetized ions exceeds the local ion-acoustic speed \cite{oppenheim_kinetic_2013}. The requirement for a large relative drift means that the structures are found in Hall currents regions, the auroral electrojets. Radio signals that scatter off these waves have been coined the `radar aurora' \cite{hysellRadarAurora2015}.

The ionosphere largely draws energy from diffuse aurorae at night \cite{newellSeasonalVariationsDiffuse2010,robinsonCalculatingIonosphericConductances1987}, produced when hot electrons near the ring current interact with naturally occurring waves through cyclotron resonance, causing pitch angle scattering and subsequent precipitation into the atmosphere \cite{ni_resonant_2008,birn_particle_2012,ni_chorus_2014}. Mechanisms  notably include  whistler-mode chorus and electrostatic cyclotron harmonic (ECH) waves \cite{horne_diffuse_2003,ni_resonant_2008,liang_themis_2010,ni_efficient_2012,kurita_observational_2014}.

Recent advances have demonstrated that temporal oscillations in chorus wave activity are closely linked to the dynamic behaviour of pulsating aurorae  \cite{kasaharaPulsatingAuroraElectron2018,hosokawaMultipleTimescaleBeats2020,miyoshi_penetration_2021}. This important discovery documents that the signature of \emph{magnetospheric} processes can be reproduced  in \emph{ionospheric} processes.

The physical route for such processes to impact the ionosphere is twofold. First, the precipitating particles quasi-instantaneously introduce a perpendicular electric field. Second, 
local ionization of the plasma  will modulate conductivities through  chemistry with its fast increase in ion production rates courtesy of impacting particles recombining with the surrounding plasma, albeit somewhat more slowly \cite{schunkTheoreticalStudyLifetime1987}.

Conjugate radar observations have established a firm link between auroral pulsations and electrodynamic oscillations in the ionosphere. Notably, periodic modulations in the plasma density, conductivity, and electric field strength are associated with pulsating aurorae \cite{hosokawaModulationIonosphericConductance2010,hosokawaPlasmaIrregularitiesAdjacent2010}. \emph{In-situ} observations have revealed the presence of downward field-aligned (likely thermal) currents on the edges of such patches \cite{gilliesSwarmObservationsFieldaligned2015}, and the filamentation (field-tube structuring) of those currents have been found to match the spatial signature of E-region plasma turbulence \cite{ivarsenDistributionSmallScaleIrregularities2023,ivarsen_turbulence_2024-1}.

Motivated by the above, we hypothesize that turbulent structuring of the electric and density fields in the auroral E-region can be, at their core, \emph{directly} caused by the wave-particle interactions in the diffuse aurora, with the aurora as a mediator of the driving signal. In testing this hypothesis for a case study, FB turbulence take center stage. Individual FB waves in the radar aurora are so intense and short-lived that they quickly dissipate and are replaced by new waves \cite{prikryl_doppler_1988,prikryl_evidence_1990,ivarsen_deriving_2024}. When and where FB waves grow will thus depend on the spatio-temporal location of their sources.

\section{\label{sec:event}Data}

\begin{figure}
    \centering
    \includegraphics[width=0.5\textwidth]{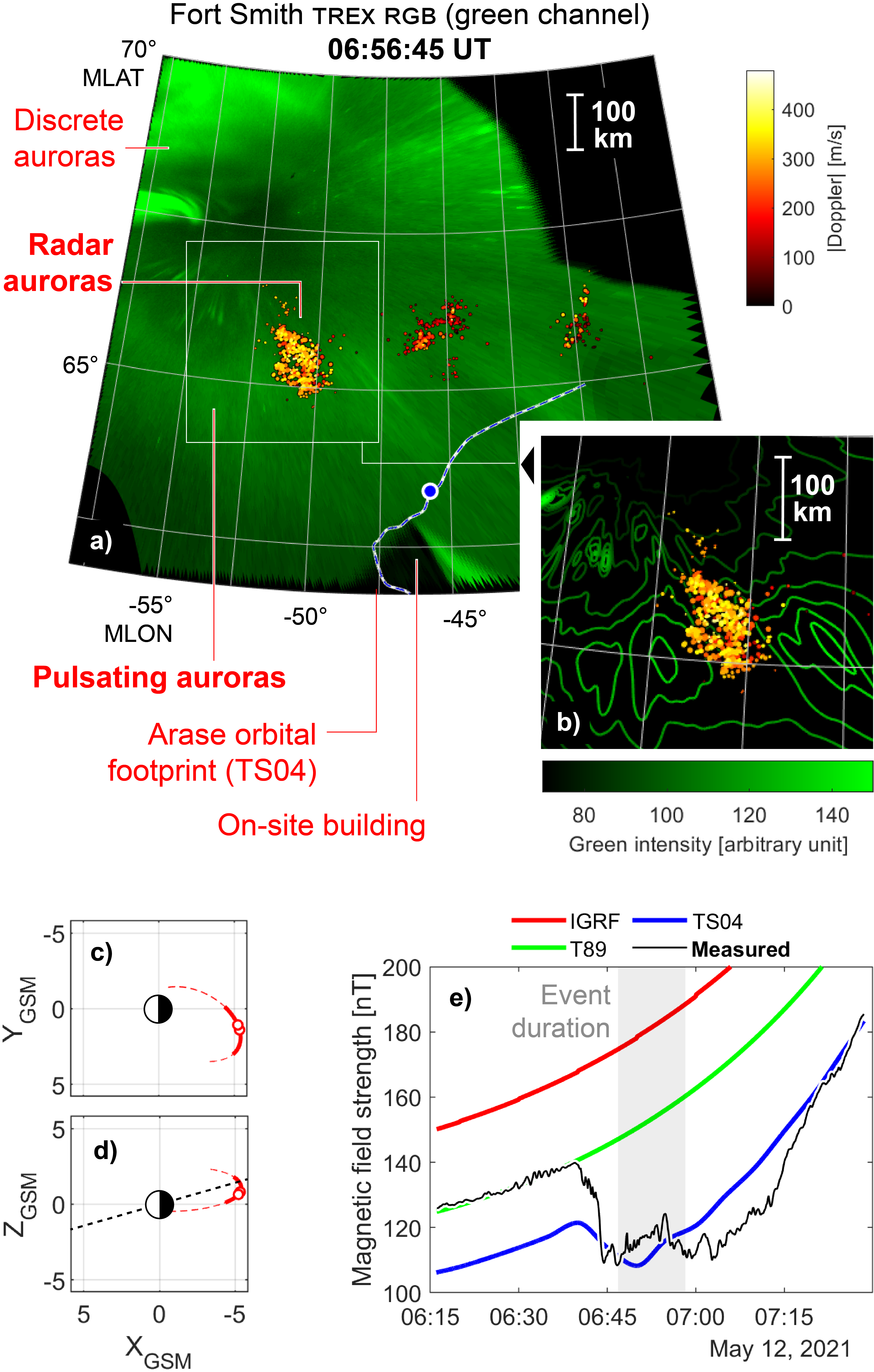}
    \caption{A conjunction on 12 May 2021 between the \textsc{icebear} radar, a \textsc{tre}x auroral camera, and the ionospheric footprint of Arase.
    \textbf{Panels a--b)} show a 3-second auroral image (the green channel of the \textsc{rgb}-triplet) projected onto the ionosphere following Ref.~\cite{gilliesApparentMotionSTEVE2020}, with the locations of radar echoes (color-coded according to Doppler shift), all shown in geomagnetic coordinates (\textsc{aacgm}, \cite{shepherd_altitude-adjusted_2014}). The ionospheric footprint of the magnetospheric spacecraft Arase is shown as a blue circle. An inset panel (b) enlarges a radar aurora shape that appears lodged between two pulsating patches (see also Video~S2).
    \textbf{Panels c--d)} detail Arase's journey in the magnetosphere in GSM coordinates, with magnetic equator appearing as a dashed black line (c--d). \textbf{Panel e)} compares the modeled (red, green, and blue) with the measured (black) magnetic field strength along Arase's trajectory. See Figure~\ref{fig:cartoon} in the End Matter for a graphical representation.}
    \label{fig:icebear}
\end{figure}

To investigate whether magnetospheric wave activity can act as a direct driver of small-scale plasma turbulence in the ionosphere, we searched for space-ground conjunctions that took place when the Japanese inner-magnetosphere spacecraft Arase was close to magnetic equator, with a northern hemisphere footprint that was  within the field-of-view of the new Canadian \textsc{icebear} radar, and in the presence of active aurorae. Combing through coincident data collected between January 2020 and June 2023, we found one such event.

Figure~\ref{fig:icebear}a) shows an annotated optical image of diffuse, pulsating aurorae, taken with the \textsc{tre}x \textsc{rgb} system at Rabbit Lake \cite{gilliesFirstObservationsTREx2019}. The system was switched on at dusk, around midway into the event. Superposed on Figure~\ref{fig:icebear}a) is the point-cloud distribution of a few thousand radar echoes that were detected in a 3-second interval following 06:56:45~UT on 12 May 2021. The radar echoes were measured by \textsc{icebear}, a  coherent scatter radar that combines multiple interferometry links with a coded pseudo-random continuous-wave signal to achieve high resolution 3D radar data\cite{huyghebaertICEBEARAlldigitalBistatic2019,lozinskyICEBEAR3DLowElevation2022}.

In Figure~\ref{fig:icebear}a), and in Videos~S1 and S2 in the supplementary data, the radar echoes cluster between evolving pulsating auroral patches along their poleward flank. Electric field enhancements maximize \textit{outside} the patches \cite{hosokawaPlasmaIrregularitiesAdjacent2010}, as opposed to the patches' interiors, where elevated conductivities may short out the field entirely \cite{maynardExampleAnticorrelationAuroral1973,gilliesSwarmObservationsFieldaligned2015}. Each received echo indicates the presence of turbulent electrojet currents in the space around the pulsating patches.

The direct cause of these turbulent currents were observed by Arase in the distant equatorial magnetosphere. Arase's orbit during the event is shown in Figure~\ref{fig:icebear}c--d). The orbit has an apogee of 32,000~km and a perigee of 400~km, an inclination angle of 31$^\circ$, and a period of 570~minutes \cite{miyoshi_geospace_2018}. Figure~\ref{fig:icebear}e) substantiates the relative accuracy of the TS04 model \cite{tsyganenko_modeling_2005} in mapping Arase's magnetic footprint from its orbit down to the northern hemisphere E-region, a mapping that is in general made uncertain by the ubiquitous presence of Alfv\'{e}n waves and field-aligned currents \cite{borovsky_dc_2001}. Arase's footprint is estimated to lie on the equatorward flank of the pulsating aurorae, whose poleward flank is occupied by radar echoes.

\begin{figure}
    \centering
    \includegraphics[width=.495\textwidth]{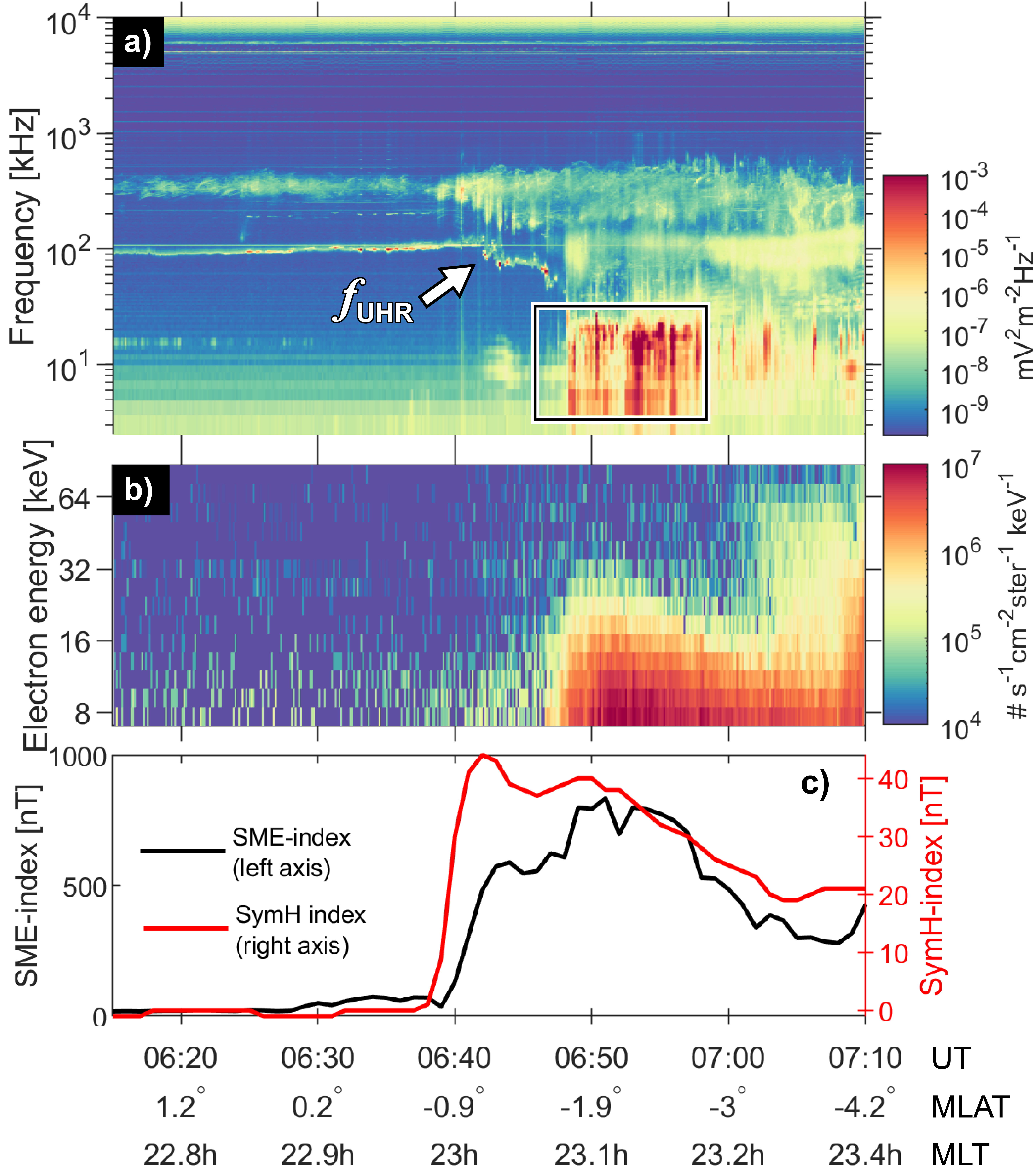}
    \caption{
    \textbf{Panel a):} High-frequency electric field spectrogram from Arase's PWE instrument. The black rectangle refers to the analysis in Figure~\ref{fig:20210512_2}). \textbf{Panel b):} precipitating electron energy flux from Arase's medium-energy particle detector (data from the low-energy detector was unavailable). The data shows pitch angles lower than 10$^\circ$, but we additionally confirmed that it was consistent with the parallel-flux data (pitch angles $<2^\circ$).  \textbf{Panel d):} The SME-index (left axis) and the SymH-index (right axis).  The three $x$-axes show time in UT (on 12 May 2021), and Arase's orbital location in the magnetosphere.}
    \label{fig:events}
\end{figure}

The magnetospheric measurements used in the present letter are summarized in Figure~\ref{fig:events}. We analyze mostly  electric field power  spectra on frequencies between 0.1~kHz and 20~kHz, a range in which whistler-mode chorus waves and electrostatic cyclotron harmonic waves often appear. The measurements originate in the Plasma wave experiment (PWE) instrument onboard Arase. The data  product consists of electric field power spectrograms \cite{kasahara_plasma_2018,kumamoto_high_2018}, shown in Figure~\ref{fig:events}a). In addition to wave power, we also collected data from Arase's onboard  Medium-energy particle detector (MEP-e), which detects  7~keV to 87~keV precipitating electrons \cite{kasahara_medium-energy_2018}, shown in Figure~\ref{fig:events}b). Lastly, Figure~\ref{fig:events}c) shows the auroral electrojet index (SME-index, black line, left axis), and the SymH-index (red line, right axis).

\subsection{Geospace description}

We are now in a position to formulate and interpret the conditions of geospace during our event. The conjunction, which was preceded by a geomagnetically quiet period, took place during drastic and coincident increases in the SME- and SymH-indices (Figure~\ref{sec:event}c). The strong impulse in these indices was the effect of a sudden nine-fold jump in the solar wind dynamic pressure (not shown), which led to magnetospheric compression \cite{shi_magnetosphere_2020}. 

Around 06:40~UT, Arase, situated at the equator ($-1^\circ$~MLAT), suddenly found itself outside the plasmasphere, witnessed by the rapid decrease in the upper hybrid resonance frequency ($f_{\text{UHR}}$ in Figure~\ref{fig:events}a, \cite{miyoshi_rebuilding_2003}). Arase was thus experiencing optimal conditions for the observation of ECH waves \cite{meredith_survey_2009,ni_resonant_2011}, waves that were simultaneously given access to hot ring current electrons \cite{ahnAuroralEnergyDeposition1989}. The loss of these electrons into Earth's atmosphere resulted in the diffuse aurorae observed with the \textsc{tre}x \textsc{rgb} system in Figure~\ref{fig:icebear}a--b) and in Video~S2.

\subsection*{Penetrating electric fields}

Next, we shall compare the wave power  measured by Arase with the FB turbulence echo \emph{detection rates}  measured by the \textsc{icebear} radar, the rate at which \textsc{icebear}'s range and Doppler gates are receiving signals. That metric is exceedingly simple: it entails counting the raw number of radar echoes detected for each timestep (1~s). With the premise of extant FB waves being ephemeral, changes in this metric will reflect changes in the turbulent driver within \textsc{icebear}'s field-of-view.

\begin{figure}
    \centering
    \includegraphics[width=.5\textwidth]{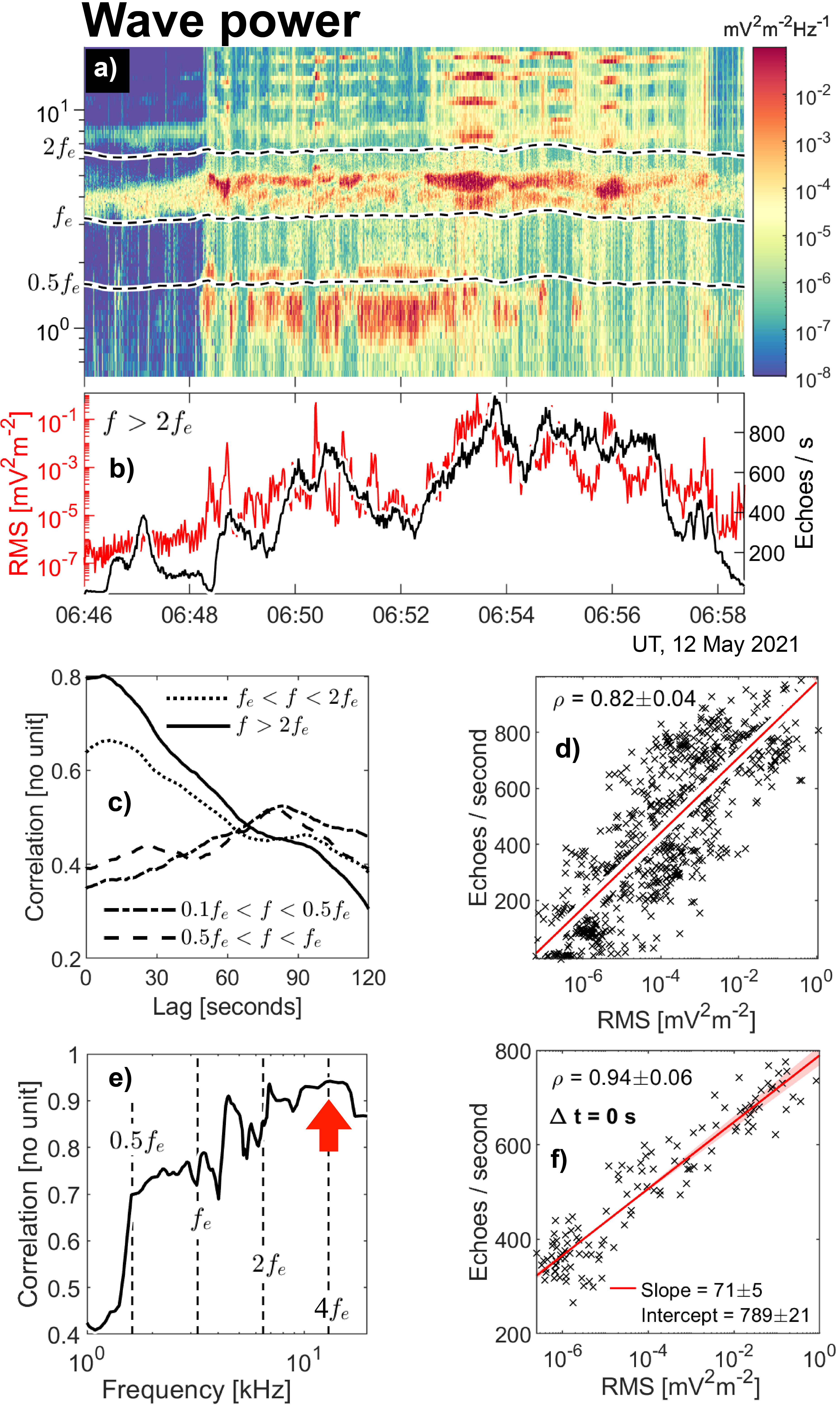}
    \caption{\textbf{panel a):} Electric field spectrogram observed between 06:46:00~UT and 06:58:30~UT on 12 May 2021, with three frequencies indicated ($0.5f_e$, $f_e$, and $2f_e$). \textbf{Panel b)}: RMS for the $2f_e<f<19.45$~kHz frequency range, with the \textsc{icebear} echo detection rate superposed (right axes). \textbf{Panel c):} cross-correlation analysis between the echo rates and the RMS for all four frequency ranges. \textbf{Panel d):} scatterplot of the data in panel b), with the ECH wave power having been shifted 7~seconds back in time; Pearson correlation coefficient for a log-linear fit, with an error margin given by 95-percent (3-sigma) confidence intervals is indicated. \textbf{Panel e):} the highest cross-correlation obtained from a moving window in frequency (7 logarithmic increments out of 132) and time (2~minutes), with a maximum allowed lag of 9~seconds. \textbf{Panel f):} scatterplot akin to Panel~d), but for frequencies around $4f_e$ (11.5~kHz$<f<$15~kHz), during a 2-minute window centered on 06:52:32~UT (at zero lag).}
    \label{fig:20210512_2}
\end{figure}

Figure~\ref{fig:20210512_2}a) shows wave power at 1~second cadence \cite{matsuda_onboard_2018}, encompassing the interval 06:46:00~UT -- 06:58:30~UT, and highlighting three frequencies   with dashed black lines ($0.5f_e$, $f_e$, and $2f_e$, with $f_e$ being the local electron cyclotron frequency, calculated using the DC magnetic field data from the MGF instrument \cite{matsuoka_arase_2018}). Figure~\ref{fig:20210512_2}b) uses a solid red line to show the integrated power (referred to as RMS, or root-mean-square) in the $f>2f_e$ range, and  we now superpose the \textsc{icebear} echo detection rate with a solid black line (right axis). Figure~\ref{fig:20210512_2}c) shows the result of a cross-correlation analysis performed between the echo detection rates and the wave power RMS in four frequency ranges. For the $f>2f_e$ range, we observe strong correlation (Pearson coefficient $\rho=0.82$, shown in Panel d), with a peak lag of 7~seconds, indicating that optimal correlation is obtained by shifting the Arase-observations 7~seconds back in time. This lag should be compared to the time-of-flight for the electrons of around 1~s \cite{miyoshiTimeFlightAnalysis2010,fukizawa_electrostatic_2018}, and we note that the correlation is similar for a 1~s lag. Figure~\ref{fig:20210512_2}e) shows explicitly how this result depends on frequency. It shows a cross-correlation analysis on a moving window both in time and frequency. There is a noticeable uptick in correlation at $f=2f_e$, and at $f=4f_e$ the correlation coefficient reaches 0.94. We note that this latter correlation was determined at a zero-lag ($<1$~second), on the order of the time-of-flight of the electrons themselves. The $p$-values (probabilities that the correlations were spurious) obtained for the two quoted coefficients were zero at floating point precision.

\subsection*{Ionization in the Lower Ionosphere}

Having demonstrated a close correspondence between the evolution of ECH wave power and that of turbulence echo detection rate, we shall next demonstrate a link between the evolution of the particle flux at Arase's orbit, and that of the echo detection \emph{altitudes}. Figure~\ref{fig:pp}a) plots the 8~keV--87~keV-electron energy flux. We then assume the flux is a proxy for the real precipitating energy flux, treating it to be caused by 16 mono-energetic beams of electrons (one for each energy channel). We apply parameterizations to estimate ionization altitude profile for each beam of electrons. The cumulative profile then corresponds to the emission altitude profile of the observed aurorae.

The results of this altitude analysis are shown in Figure~\ref{fig:pp}c--d), where we likewise plot the total altitude distribution of the radar echo point cloud. The two timeseries in Figure~\ref{fig:pp}c) exhibit a Pearson correlation coefficient of 0.87, and the two profiles in Figure~\ref{fig:pp}d) match surprisingly well, with peaks in the distributions being separated by only 1.5~km. The unambiguous similarity points to a causal link between the particle-induced ionization and the distribution of radar echoes in the E-region.


\section{Discussion}

At the event's onset, Arase had just exited the plasmasphere at the equator (orbiting from $-1.5^\circ$ to $-3^\circ$~MLAT) where it observed intense ($0.1$ mV$^{-2}$m$^{-1}$) ECH waves inside the region where such waves are confined \cite{meredith_survey_2009,ni_resonant_2011}, and capable of accelerating keV-electrons towards Earth's atmosphere \cite{horne_diffuse_2003,meredith_survey_2009,kurita_observational_2014,fukizawa_statistical_2022}.

\begin{figure}
    \centering
    \includegraphics[width=0.5\textwidth]{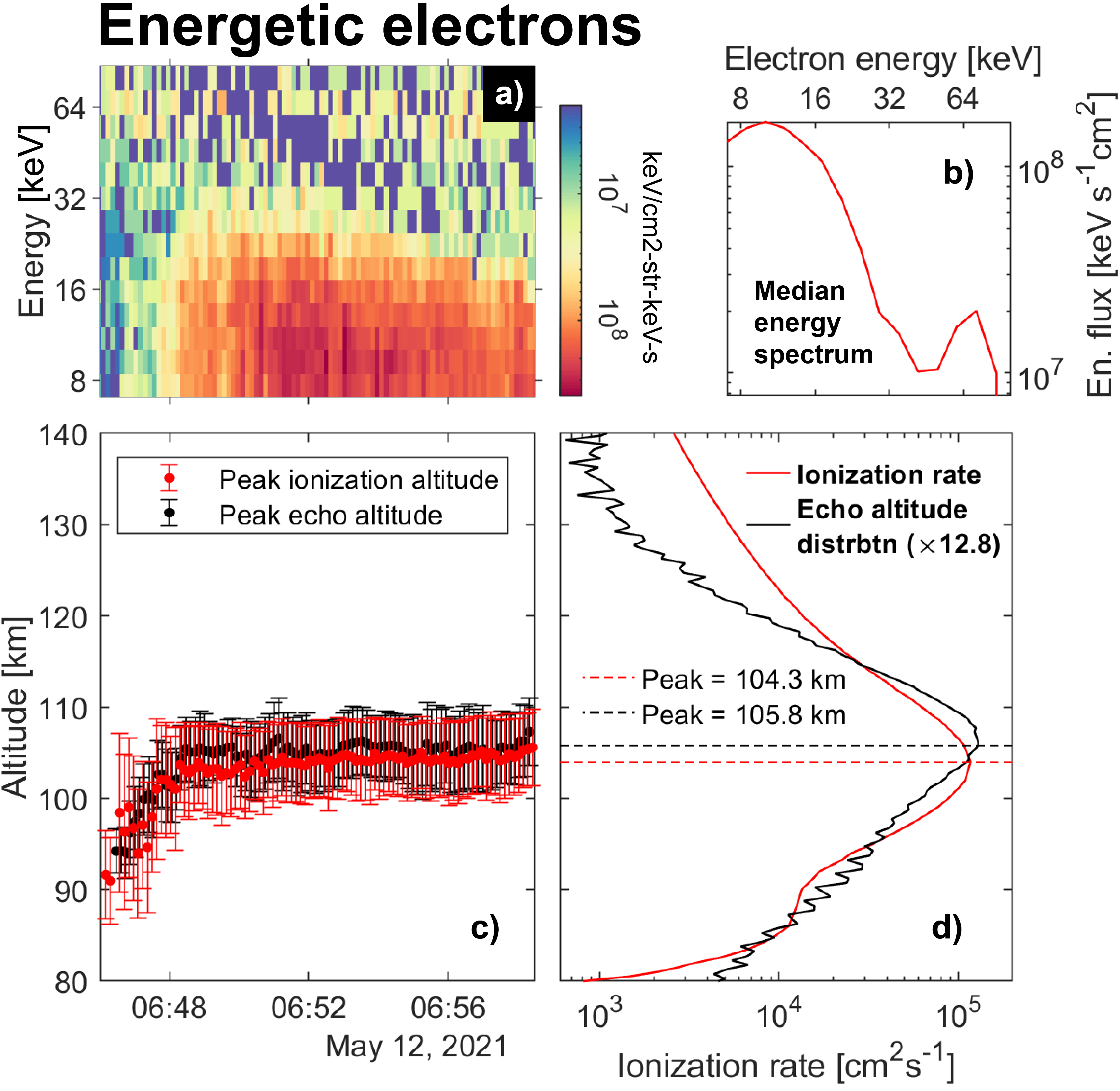}
    \caption{\textbf{Panel a):} Precipitating particle energy flux measured by Arase. \textbf{Panel b):} Median energy spectrum. \textbf{Panel c):} \textsc{icebear} echo altitudes as a function of time (black dots) and the peak emission altitude based on the electron energy flux (red errorbars), using generalized parametrization of numerical models \cite{fangParameterizationMonoenergeticElectron2010}, including the MSIS model of Earth’s atmosphere \cite{piconeNRLMSISE00EmpiricalModel2002}. \textbf{Panel d):} overall \textsc{icebear} echo altitude distribution (black) and  auroral emission altitude profile (red), peak altitudes indicated.}
    \vspace{-15pt}
    \label{fig:pp}
\end{figure}

Arase's footprint in the ionosphere was likely 100 -- 200~kilometers south of the echo region, with diffuse auroral shapes appearing in between. The precipitating electrons inside the patches briefly accumulated at an altitude of around 104~km. There, they produced strong electric fields \cite{opgenoorthRegionsStronglyEnhanced1990,lanchester_relationship_1996} and plasma density gradients by merit of ionizing the gas. The plasma can then become gradient-drift unstable  \cite{greenwaldDiffuseRadarAurora1974}, and at the same time, the Farley-Buneman instability may saturate \cite{schmidt_density_1973}. Once saturated the instability produces secondary waves  \cite{oppenheim_saturation_1996,otani_saturation_1998} and dissipate the wave power through heating \cite{st-mauriceElectronHeatingPlasma1990,st-mauriceRevisitingBehaviorERegion2021}. The high-amplitude FB waves dissipate fast, giving the turbulence the impression of being ephemeral, or instantaneous,  \cite{ivarsen_deriving_2024}. Ultimately, this allows the ensemble radio echoes from those waves to accurately reflect the changes in the instability driver on timescales larger than a second, facilitating the strong correlations between the various timeseries in Figure~\ref{fig:20210512_2}d--f) and Figure~\ref{fig:pp}c).

The resulting image is one of a self-similar and turbulent state, simultaneously measured at points in space separated by 5 Earth radii. It implies that the entire turbulent process retains a distinct `stochastic shape' that is conserved down to meter-scale.

There is a causal and largely instantaneous path from the modulation of ECH waves to small-scale turbulence. It involves particle precipitation, which creates beams of electrons that modulate conductivity (ionization) and electric fields \cite{hosokawaElectricFieldModulation2008}. Those modulations can produce gradient-drift unstable structures \cite{greenwaldDiffuseRadarAurora1974}, whose imminent decay into smaller and smaller pieces, the turbulent cascade, systematically breaks apart structures smaller in size than around 1--10~km  \cite{ivarsenLifetimesPlasmaStructures2021,ivarsen_plasma_2024}. The shapes, or rather, their stochastic features, are subsequently repeated in a self-similar pattern down to meter-scale in magnification, at which point the initial perturbations can seed the Farley-Buneman instability and saturate its growth rates (see Figure~\ref{fig:cartoon} in the End Matter). 

Other driving mechanisms are readily available, such as Alfv\'{e}n waves and other field-parallel acceleration mechanisms, as well as magnetospheric ion cyclotron waves, and other wave-particle interactions. However, as shown by Figure~\ref{fig:20210512_2}d--f), the ensemble of radar echoes recreate temporal changes in the magnetospheric wave-particle interactions with remarkable fidelity: the production of small-scale plasma turbulence identifies its maker, faithfully mimicking its driving signal, ostensibly by \textit{dissipating} that signal. 

Refreshingly and for a brief moment, then, the local space weather had a clear physical driver. The turbulent transformation of this driving signal can inform development of models that aim to predict the occurrence of plasma turbulence around aurorae.

\section{Summary}

In this letter, we have reported a conjunction between Arase's northern hemisphere footprint and the \textsc{icebear} radar's field-of-view. Arase observed strong electrostatic cyclotron harmonic wave activity at the magnetic equator. On a nearby magnetic field line, \textsc{icebear} recorded a matching radar signal from turbulent electrojets in the space between pulsating aurorae. 

Our interpretation of the findings, and the only viable explanation we can find for their cause, is that an ensemble of wave-particle interactions imparted their temporal and spatial characteristics to a spatio-temporal pattern of electron precipitation. The electrons inside these structures produced pulsating aurorae and introduced an associated pattern of ionization and electric field enhancements in the E-region that were insensibly picked up by the growth of FB waves.   The growth, saturation, and subsequent detection of these waves faithfully reflected the evolution of the underlying driver. 


We conclude that the detection of small-scale plasma turbulence in the auroral E-region can be applied as a diagnostic tool to quantify the whole energy input behind space weather events, a remarkable victory for the concept of mode-coupling in plasma physics. 





\balancecolsandclearpage

\section*{End Matter}

\subsection{A Graphic Representation of the Findings}

\begin{figure}
    \centering
    \includegraphics[width=0.5\textwidth]{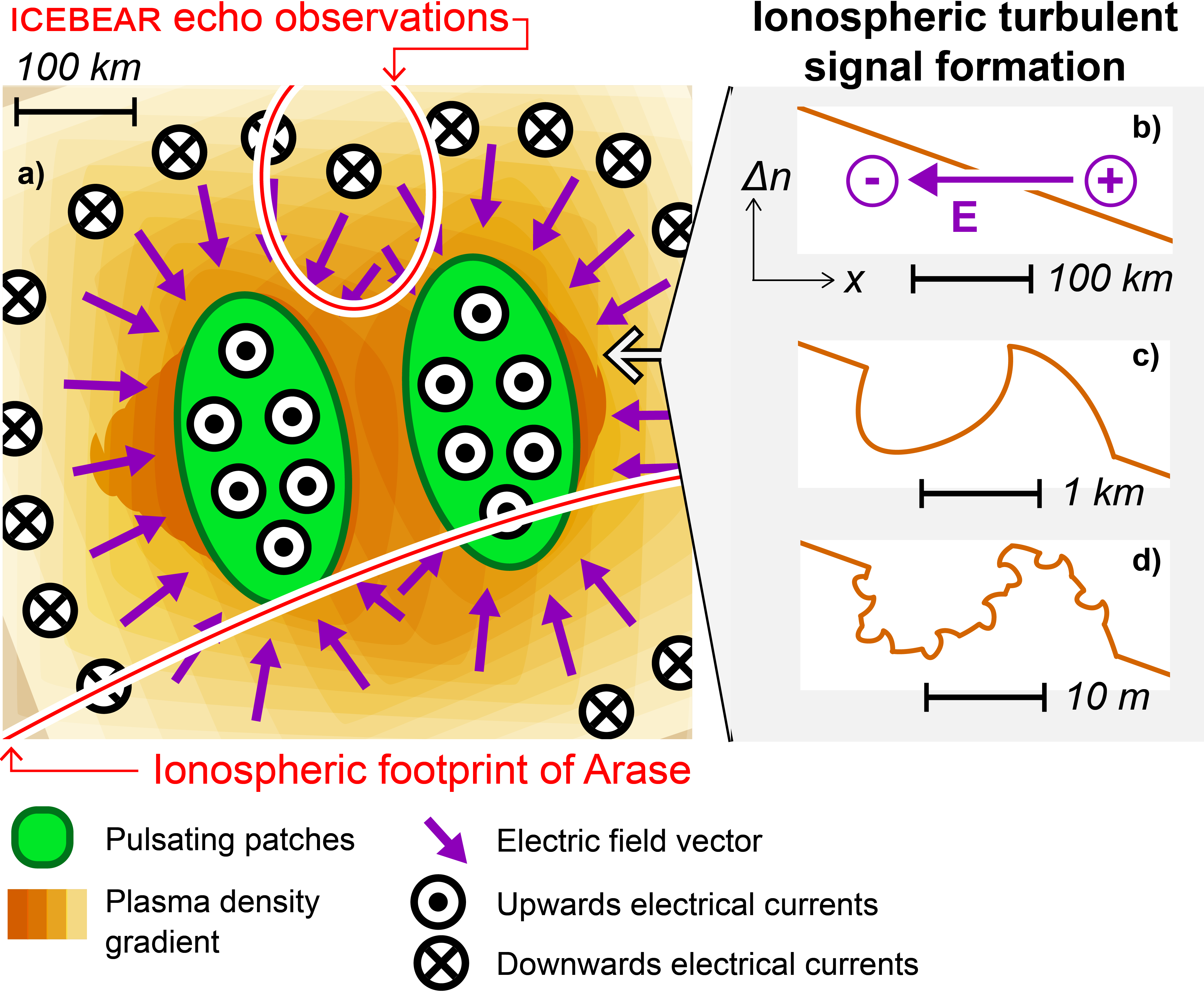}
    \caption{\textbf{Panel a):} Annotated schematic depiction of the electrodynamics surrounding the \textsc{icebear} and Arase measurements. \textbf{Panel b):} A schematic representation of a density structure near one of the pulsating patches in panel a), along the density gradient ($\Delta n$), and with an applied electric field ($\boldsymbol{E}$). Together these may generate sub-kilometer-scale turbulent structures through the gradient drift instability. The irregular structures can further decay into smaller-and-smaller pieces, until they may seed the meter-scale Farley-Buneman instability, itself likewise triggered by $\boldsymbol{E}$. Panels b--d) are each $100\times$ magnifications of the foregoing panel.}
    \label{fig:cartoon}
\end{figure}

Figure~\ref{fig:cartoon}a) illustrates the electrodynamics of the conjugate measurements analyzed in the present letter. Intense electric fields, which point in the direction of moving ions, are organized in some pattern outside of pulsating auroral patches \cite{hosokawaPlasmaIrregularitiesAdjacent2010}. The field points away from downwards (likely thermal) currents and towards the upwards currents (aurorae). Surrounding the pulsating patches are plasma density gradients \cite{hosokawaModulationIonosphericConductance2010}. 

Figure~\ref{fig:cartoon}b--d) illustrate how this description can turn into turbulent energy dissipation, as the density structures in Panel a) become gradient-drift unstable \cite{greenwaldDiffuseRadarAurora1974,tsunodaHighlatitudeRegionIrregularities1988}. The resulting waves decay into smaller-and-smaller turbulent branches, until they may act as seeds for the meter-scale Farley-Buneman instability. Thus, a turbulent signal is formed by \textit{the ensemble} of 3-meter Farley-Buneman waves, one whose temporal characteristics approach that of the wave power, the underlying driver.

The approximate measurement locations of those two matching turbulent signals are indicated around the pulsating patches in the electrodynamic diagram in Figure~\ref{fig:cartoon}a), and Figure~\ref{fig:cartoon}b--c) illustrate the turbulent signal formation in the lower ionosphere.

\subsection{Part of a Trend?}

We argue that the clear-cut event analyzed in the present letter is part of a trend, where Ref.~\cite{ivarsen_transient_2025} presents relevant evidence for this trend from the dayside polar ionosphere. Indeed, the turbulent Hall channels that may frequent the vicinity of diffuse aurorae in general are highly localized in time and space, like most research into the ``spiky" nature of electric field enhancements can attest to \cite{opgenoorthRegionsStronglyEnhanced1990}. To support such a general connection between magnetospheric wave-particle energy and subsequent modulations to the \textsc{icebear} echo detection rates, we have fallen back on inferences made from statistics.

We aggregated the echo detection rates between magnetic local times of 20h and 06h, a distribution that is roughly consistent with surveys of ECH waves \cite{ni_resonant_2011,meredith_survey_2009}. We took the raw number of echoes per second after removing bins with a mean signal-to-noise ratio lower than 1.5, which removes the occasional radio interference, and discarding bins in which only a single echo was detected. We subsequently aggregated Arase ECH wave power RMS for frequencies $f>2f_e$, when the satellite was within 5$^\circ$ of the equator at the same magnetic local time interval, and with a northern hemisphere orbital footprint between $60^\circ$ and $72^\circ$~MLAT. 

\begin{figure}[b!]
    \centering
    \includegraphics[width=.495\textwidth]{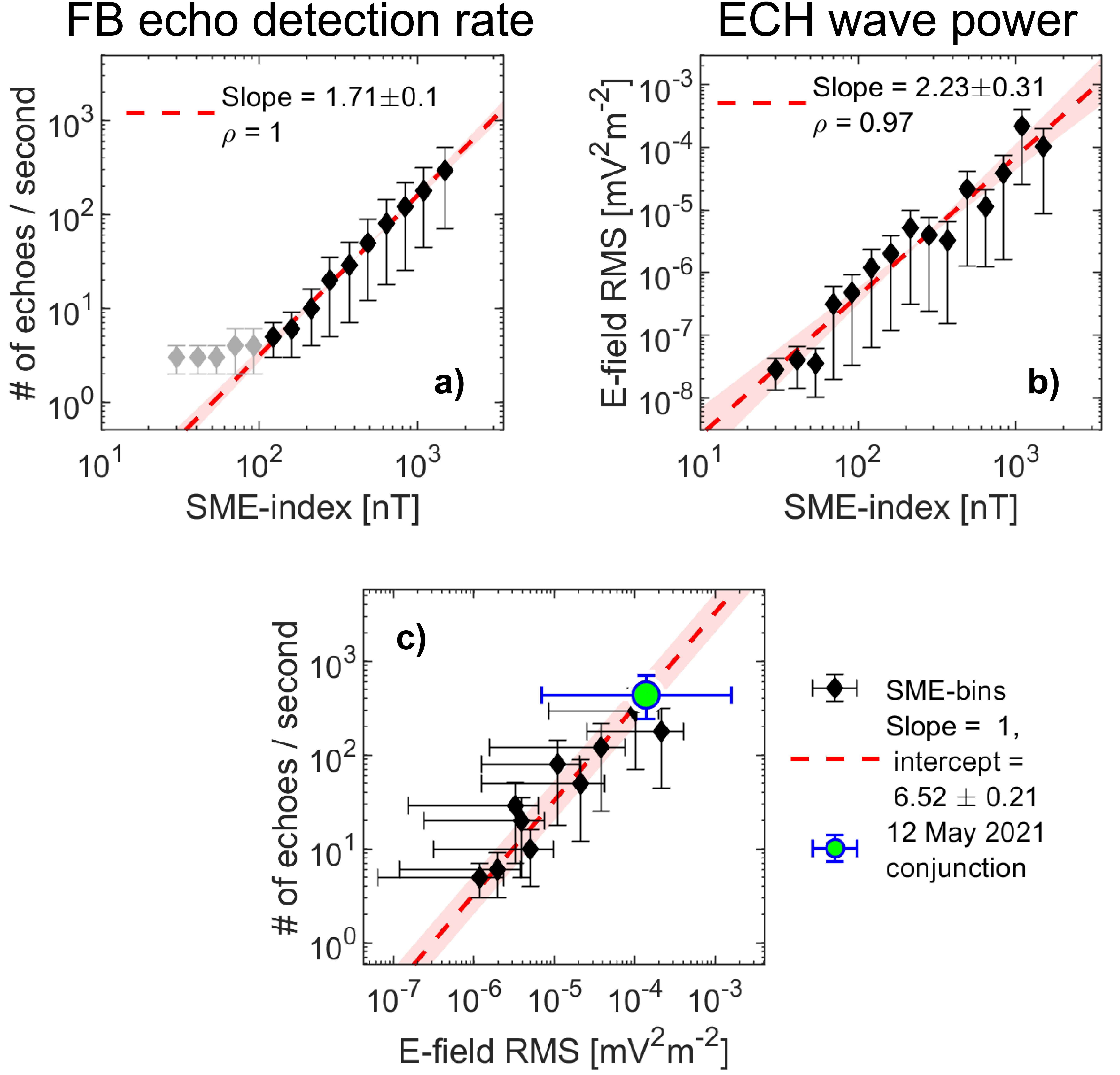}
    \caption{The \textsc{icebear} echo detection rate, for some 1.2~million one-second intervals (containing a total of 272~million echoes) \textbf{(Panel a)}, and the Arase-observed wave  RMS ($f>2f_e$) for some 530,000 one-second wave spectra \textbf{(Panel b)}. Black diamonds represent the median value of 15 logarithmically spaced SME-index bins, and vertical errorbars denote upper/lower quartile distributions. Linear fits are indicated, by non-linear least squares minimization of the root-mean-square error, and error margins are posted showing 95-percent confidence intervals of the fits (3-sigma). \textbf{Panel c)} shows the same bins in a scatterplot, with a red dashed line denoting a 1:1 relation. The data was collected between January 2020 and June 2023 during days when the radar was operational. Data from the 12 May 2021 event is shown with a green circle. 
    } 
    \label{fig:stats}
\end{figure}

In Figure~\ref{fig:stats}a--b) we show the resulting median values for the echo detection rates and wave power respectively, in 15 geomagnetic activity bins. In Panel a) we observe a distinct kink in the curve at an SME-index value of around 150~nT, below which the echo detection rate is flat (indicating a contamination by meteor trail echoes \cite{ivarsenAlgorithmSeparateIonospheric2023}). For bins above this kink, the bin-median echo detection rate is perfectly correlated ($\rho=1$) with the SME-index. The linear slope ($1.71\pm0.11$ in a log-log scale) is roughly consistent with the slope exhibited by the wave power data (Figure~\ref{fig:stats}b), $2.23\pm0.31$.

The two quantities exhibit a similar response to enhancements in the SME-index. This facilitates the 1:1 relationship shown in Figure~\ref{fig:stats}c), where we display the same geomagnetic activity bins in a scatterplot (excluding the gray datapoints in Panel a as well as the equivalent bins in panel b), and a log-log linear fit with slope 1 is shown with a red, dashed line. Lastly, with a green-blue circle, we show the median values derived from the 12 May 2021 event, with errorbars denoting upper/lower quartile distributions.

Figure~\ref{fig:stats}a-b) show two widely different geophysical quantities -- one in the ionosphere and one in the magnetosphere -- that nevertheless respond in a similar fashion to a \emph{third} independent variable, the SME-index. Indeed, that index is itself an unambiguous measurement from the ground of the ionosphere's high-latitude Hall currents (the auroral electrojets) \cite{davisAuroralElectrojetActivity1966,baumjohann03MagnetosphericContributions2007}. Enhancements in Hall conductance are, in turn, driven by energetic particle precipitation  \cite{vickreyDiurnalLatitudinalVariation1981,coumansGlobalAuroralConductance2004}, by virtue of providing ionization and causing electric field enhancements. As we detail in the present letter (and illustrate in Figure~\ref{fig:cartoon}), the same two quantities are able to drive Farley-Buneman turbulence.

In empirical terms, the diffuse and pulsating aurorae constitute the majority of the total energy input into the nightside ionosphere \cite{newellDiffuseMonoenergeticBroadband2009}, and so enhancements in  the Hall currents there are largely driven by diffuse and pulsating aurorae \cite{hosokawaModulationIonosphericConductance2010,blandDregionImpactArea2021}. 

The \textsc{icebear} echo rate reflects the amplitude and spatial extent of any large-amplitude meter-scale Farley-Buneman-generated  turbulence in the electrojets within the radar field-of-view \cite{sahr_auroral_1996,oppenheim_large-scale_2008,fejerIonosphericIrregularities1980}.

The above chain of argument explains why $\rho=1$ in Figure~\ref{fig:stats}a): 
Global though it may be, and thus not expected to correlate with the echo rates during individual events, the SME-index encapsulates the average response of the \textsc{icebear} data in the face of an externally driven magnetosphere.



\section*{Acknowledgements}
This work is supported in part by Research Council of Norway (RCN) grant [324859]. We acknowledge the support of the Canadian Space Agency (CSA) [20SUGOICEB], the Canada Foundation for Innovation (CFI) John R. Evans Leaders Fund [32117], the Natural Science and Engineering Research Council (NSERC), the Discovery grants program [RGPIN-2019-19135], the Digital Research Alliance of Canada [RRG-4802], and basic research funding from Korea Astronomy and Space Science Institute [KASI2024185002]. Science data of the ERG (Arase) satellite were obtained from the ERG Science Center operated by ISAS/JAXA and ISEE/Nagoya University (\url{https://ergsc.isee.nagoya-u.ac.jp/index.shtml.en}. This includes Lv.3 MEP-e (DOI \texttt{10.34515/DATA.ERG-02003}), Lv.2 PWE/OFA (DOI \texttt{10.34515/DATA.ERG-08000}), Lv.2 PWE/HFA (DOI \texttt{10.34515/DATA.ERG-10000}), and Lv.2 MGF (DOI \texttt{10.34515/DATA.ERG-06001}). \textsc{icebear} 3D echo data for 2020, 2021 is published with DOI \texttt{10.5281/zenodo.7509022}. Super\textsc{mag} data can be accessed at \texttt{https://supermag.jhuapl.edu/mag/}. The SymH-index from NASA's \textsc{omni} service can be accessed at \url{https://omniweb.gsfc.nasa.gov/}. \textsc{tre}x optical data can be accessed at \url{https://doi.org/10.11575/4P8E-1K65}. MFI is thankful to F Lind, Y Jin, W Miloch, and L Clausen for stimulating discussions.
\\
\\
See Supplemental Material [url] for additional frequency-dependent correlation akin to Figure~\ref{fig:20210512_2}, as well as a brief discussion of lower-band chorus waves in the same context. The Supplemental Materials includes Refs.~\cite{li_global_2009,spasojevic_drivers_2010,aryanGlobalMapChorus2022,troyerSubstormDrivenChorus2024,hosokawaPedersenCurrentCarried2010}.


%

\end{document}